\IfFileExists{ptephy_v1.cls}{%
  \documentclass[preprint]{ptephy_v1}
}{%
  \documentclass[11pt]{article}
  \usepackage[a4paper,margin=25mm]{geometry}
}

\usepackage{amsmath,amssymb,amsfonts}
\usepackage{bm}
\usepackage{booktabs}
\usepackage{hyperref}
\providecommand{\preprintnumber}[1]{}
\providecommand{\affil}[1]{\date{#1}}
\providecommand{\email}[1]{\\\texttt{#1}}
\providecommand{\subjectindex}[1]{}

\newcommand{\cycles}{C}
\newcommand{\cont}{c}
\newcommand{\dd}{\mathrm d}
\newcommand{\e}{\mathrm e}
\newcommand{\tri}{\mathrm{tri}}
\newcommand{\hex}{\mathrm{hex}}
\newcommand{\Dc}{D_{\rm c}}
\newcommand{\Zcyc}{\mathcal Z_Q}
\newcommand{\Ytri}{\mathcal Y_Q}
\newcommand{\Sseries}{\mathcal S}

\preprintnumber{}

\title{Star-triangle duality estimates for triangular and honeycomb
permutation models}
\author{Masayuki Ohzeki}
\affil{Department of Physics, Institute of Science Tokyo, Tokyo, 152-8551,
Japan\\
Graduate School of Information Sciences, Tohoku University, Miyagi 980-8579,
Japan\\
Research and Education Institute for Semiconductors and Informatics,
Kumamoto University, Kumamoto 860-8555, Japan\\
Sigma-i Co., Ltd., Tokyo, 108-0075, Japan
\email{mohzeki@tohoku.ac.jp}}

\begin{document}

\begin{abstract}
We study a duality analysis in conjunction with the star-triangle transformation for symmetric-group permutation models on the triangular and honeycomb lattices.  The calculation is motivated by the permutation-model description of random tensor networks and by earlier duality analyses of replicated spin glasses.  
The essential point is that the finite-basis unit is not a bare bond but a star-triangle block.
Our analysis estimates the critical bond dimension for the honeycomb lattice to be 2.634929344884, and the associated single-bond duality relation yields the triangular-lattice estimate 1.475661534848.  
\end{abstract}
\subjectindex{A-41}

\maketitle

\section{Introduction}

Duality remains one of the most efficient analytic ideas in two-dimensional
statistical mechanics.  Its archetypal use is the Kramers-Wannier argument
for the Ising model, where a high-temperature expansion is related to a
low-temperature expansion on the dual lattice and, under the usual uniqueness
assumption, the self-dual point identifies the critical temperature
\cite{Kramers1941}.  The same idea was later formulated in terms of Fourier
transformations for \(Z_q\) and many-component spin systems
\cite{WuWang1976}.  It has also been used extensively for Potts, Villain,
gauge, and related lattice models \cite{Kogut1979}.

The method is especially useful, and also subtle, in disordered systems.  
For random-bond Ising and Potts spin glasses, the replica method,
gauge symmetry on the Nishimori line, and duality lead to a scalar condition
on the principal Boltzmann factor
\cite{Nishimori1979,Nishimori2002,Maillard2003}.
This condition is highly predictive but is not, by itself, an exact
renormalized fixed-point equation.  
The finite-basis improvement introduced in later work replaces the single-bond condition by a condition on a
renormalized cluster or graph polynomial, producing rapidly convergent
estimates of multicritical points
\cite{Nishimori2006,Ohzeki2008hl,Ohzeki2009,Ohzeki2011slope,Ohzeki2015}.

Permutation models based on the symmetric group have recently appeared in a
different but closely related setting.  Replica analyses for random
tensor networks and random quantum circuits map entanglement observables to
classical statistical models whose degrees of freedom are permutations
\cite{Zhou2019,Romain2019,Yimu2020,Jian2020,Hayden2016}.  Permutation and
non-Abelian symmetry models also have a duality-related history in lattice
statistical mechanics \cite{Drouffe1979,Drouffe1978,Buchstaber2003}.  A
duality analysis directly in the symmetric group was proposed as a way to
estimate the transition of the corresponding random tensor-network
permutation model \cite{PermutationDuality}.  
A recent numerical study on
self-dual hierarchical lattices used exact real-space renormalization group
calculations to determine finite-replica critical points and found good
agreement with the duality prediction after extrapolation toward the replica
limit \cite{Ito2026}.  
This comparison gives practical support to the view that, in the permutation model, the combination of duality and the replica or quenched limit is not intrinsically a low-accuracy procedure.  
The single-bond duality analysis therefore gives a natural starting estimate,
but the spin-glass analogy suggests that the next meaningful step is to replace the bare bond by the smallest finite basis.
For the triangular and honeycomb lattices, the appropriate finite basis is
the star-triangle block.  This is the same local geometry that underlies the
standard relation between the triangular and honeycomb Ising models and the
replicated spin-glass analysis on the two dual lattices.  In particular, the
duality analysis of non-self-dual spin-glass pairs uses a product relation
between the single-bond principal Boltzmann factors of mutually dual lattices
\cite{BaxterBook,Takeda2005,Nishimori2006,Ohzeki2009}.

The remainder of this paper starts with the definition of the permutation model and the introduction of the replica limit used to assess entanglement entropy.  
We then review the principal-factor duality idea and its star-triangle use in
replicated spin glasses.  
The application to the permutation model is then
given with the single-bond triangular-honeycomb relation separated from the
star-triangle block projections, and with the group-theoretic derivations
placed in appendices.
Finally, we discuss the physical meaning of the triangular and honeycomb estimates.

\section{Model}
\label{sec:model}

We consider a ferromagnetic permutation model on a two-dimensional graph
$G=(V,E)$.  
The site variable is a permutation
\(\sigma_i\in S_Q\), and the partition function is
\begin{equation}
        Z_G(D,Q)
        =
        \sum_{\{\sigma_i\}}
        \prod_{\langle ij\rangle\in E}
        f_D(\sigma_i\sigma_j^{-1}),
        \qquad
        f_D(g)=D^{\cycles(g)} .
        \label{eq:model}
\end{equation}
Here \(\cycles(g)\) denotes the number of cycles of \(g\).  The parameter
\(D\) is the bond dimension in the random tensor-network interpretation.  It
also plays the role of a ferromagnetic coupling, since one may write
\(D=\exp K\) and regard larger \(D\) as lower effective temperature.

The replica parameter \(Q\) is first a positive integer.  The random
tensor-network application uses analytic continuation in \(Q\), and the
replica limit used below is the coefficient linear in \(Q\) at \(Q=0\).
This continuation is the same kind of operation as in the spin-glass
duality literature: the finite-integer model supplies the algebraic
identities, while the physical estimate is obtained only after the replica
coefficient is extracted.

At \(Q=2\), the symmetric group has two elements.  If the relative
permutation is the identity it has two cycles, while the transposition has
one cycle.  Up to an irrelevant overall factor the edge weight is therefore
the ferromagnetic Ising weight with
\begin{equation}
        D=e^{2K_{\rm Ising}} .
        \label{eq:D-Ising}
\end{equation}
This finite-\(Q\) case gives a useful normalization check for the triangular
and honeycomb equations derived below.

For later use we introduce the cycle-index polynomial
\begin{equation}
        \Zcyc(a)
        =
        \sum_{g\in S_Q}a^{\cycles(g)}
        =
        \frac{\Gamma(a+Q)}{\Gamma(a)} .
        \label{eq:cycle-index}
\end{equation}
The derivative at the replica point is
\begin{equation}
        \left.
        \partial_Q\log \Zcyc(a)
        \right|_{Q=0}
        =
        \psi(a),
        \label{eq:cycle-index-derivative}
\end{equation}
where \(\psi\) is the digamma function.  The derivation and the corresponding
non-Abelian Fourier identities are summarized in
Appendix~\ref{app:symmetric-group}.

\section{Star-triangle Duality and Replicated Duality}
\label{sec:review}

For a lattice model whose edge weight can be Fourier transformed, duality
maps the partition function on a planar lattice to a related model on the
dual lattice.  In self-dual cases, the equality of the original and dual
principal Boltzmann factors is a natural candidate for the critical
condition.  In non-self-dual or replicated cases, the same condition should be
viewed more cautiously: it is a scalar projection of the duality relation,
not a proof that all interactions are fixed.

For the present permutation model, the single-bond principal-factor condition
is
\begin{equation}
        D^Q
        =
        \frac{1}{\sqrt{\Gamma(Q+1)}}
        \frac{\Gamma(D+Q)}{\Gamma(D)} .
        \label{eq:single-bond}
\end{equation}
Taking the coefficient linear in \(Q\) gives
\begin{equation}
        \log D=\psi(D)+\frac{\gamma}{2},
        \label{eq:single-bond-q0}
\end{equation}
with solution
\begin{equation}
        D_0=1.882008326950\cdots .
        \label{eq:single-bond-result}
\end{equation}
This number is the bare duality estimate of Ref.~\cite{PermutationDuality}.
The star-triangle calculation below is the triangular-honeycomb analogue of a
finite-basis correction to this bare estimate.

It is useful to recall the calculation for a \(Z_q\) spin model, because it
shows explicitly where the principal-factor equation comes from.  Let the
site variable be \(s_i\in Z_q\), and let the edge Boltzmann factor depend
only on the difference, $x_{s_i-s_j}$, where $s_i-s_j\in Z_q$.
The discrete Fourier transform gives the dual edge weights
\begin{equation}
        x_m^*
        =
        \frac{1}{\sqrt q}
        \sum_{k=0}^{q-1}
        \exp\!\left(\frac{2\pi i mk}{q}\right)x_k .
        \label{eq:zq-fourier}
\end{equation}
The principal Boltzmann factor is the weight for equal neighboring spins,
\(x_0\), and the dual principal factor is
\begin{equation}
        x_0^*
        =
        \frac{1}{\sqrt q}
        \sum_{k=0}^{q-1}x_k .
        \label{eq:zq-principal-dual}
\end{equation}
For a self-dual lattice, the scalar duality estimate is therefore
\begin{equation}
        x_0=x_0^* .
        \label{eq:zq-principal-condition}
\end{equation}
For the ferromagnetic Potts model \(x_0=e^K\) and
\(x_k=1\) for \(k\ne0\), Eq.~\eqref{eq:zq-principal-condition} gives
\(e^K=1+\sqrt q\), and the \(q=2\) case reproduces the usual Ising
duality result after translating conventions.

The same logic applies to the triangular-honeycomb pair, but the elementary
object is a star-triangle block rather than a single bond.  
On the honeycomb side, fix the three external \(Z_q\) spins of a Y block to zero
and sum the central spin.  
The original and dual principal factors are
\begin{align}
        A_{0,\hex}^{(Y)}
        &=
        \sum_{s=0}^{q-1}x_s^3,
        \label{eq:zq-y-original}\\
        A_{0,\hex}^{*(Y)}
        &=
        \frac{1}{q}
        \left(\sum_{s=0}^{q-1}x_s\right)^3 .
        \label{eq:zq-y-dual}
\end{align}
The dual one is calculated by considering the dual edge weight on a Y block.
In other words, we apply the duality transformation on the triangular lattice and obtain the dual model on the honeycomb lattice.
The summation in brackets comes from the discrete Fourier transform.

The estimate of the critical point on the honeycomb lattice is given by the principal-factor relation
\begin{equation}
        A_{0,\hex}^{(Y)}=A_{0,\hex}^{*(Y)} .
        \label{eq:zq-y-condition}
\end{equation}
Conversely, on the triangular side, the original principal factor is
\begin{equation}
        A_{0,\tri}^{(\triangle)}=x_0^3,
        \label{eq:zq-tri-original}
\end{equation}
whereas the dual triangular face contains the zero-flux constraint around
the face:
\begin{equation}
        A_{0,\tri}^{*(\triangle)}
        =
        \frac{1}{q}
        \sum_{\substack{k_1,k_2,k_3\in Z_q\\
        k_1+k_2+k_3=0}}
        x_{k_1}x_{k_2}x_{k_3}.
        \label{eq:zq-tri-dual}
\end{equation}
This equation results from the discrete Fourier transformation on the honeycomb dual lattice and the summation over the degree of freedom at the center of the Y shape.
Thus the triangular estimate is
\begin{equation}
        A_{0,\tri}^{(\triangle)}
        =
        A_{0,\tri}^{*(\triangle)} .
        \label{eq:zq-tri-condition}
\end{equation}
For \(q=2\), Eqs.~\eqref{eq:zq-y-condition} and
\eqref{eq:zq-tri-condition} give the exact honeycomb and triangular Ising
critical points.  This is the Abelian prototype of the calculation carried
out below for \(S_Q\).

Even for random systems, including replica formulations relevant to the permutation model, one first performs the same duality transformation before taking the quenched limit; this gives an estimate of the critical point.  
In addition, to reduce the computational complexity, we employ the duality relation for the mutually dual pair of lattices,
\begin{equation}
        x_0^{(1)}x_0^{(2)}
        =
        x_0^{*(1)}x_0^{*(2)} ,
        \label{eq:paired-principal-general}
\end{equation}
for the single-bond principal factors of the two dual systems
\cite{Takeda2005}.  
For the triangular and honeycomb lattices, Nishimori and Ohzeki used a duality transformation and star-triangle transformation to directly obtain estimates of the critical points for the random-bond Ising model \cite{Nishimori2006}.
They confirmed that the duality relation Eq.~\eqref{eq:paired-principal-general} is satisfied by the pair of critical points on the triangular and honeycomb lattices.

In addition, the finite-basis duality analysis improves the estimate of the critical points beyond the bare-bond condition \cite{Ohzeki2009}.  
In this sense, the duality analysis, in conjunction with the star-triangle transformation, yields a fairly good estimate even for non-self-dual models, such as the random-bond Ising model.

\section{Application to the Permutation Model}
\label{sec:application}
For the star-triangle basis, we now analyze the permutation model.  
On the honeycomb Y block,
\begin{align}
        H_{0}^{(Y)}(Q,D)
        &=
        \sum_{\sigma\in S_Q} f_D(\sigma)^3,
        \label{eq:perm-y-x0-review}\\
        H_{0}^{*}(Q,D)
        &=
        \frac{1}{Q!}
        \left(\sum_{\sigma\in S_Q}f_D(\sigma)\right)^3 .
        \label{eq:perm-y-x0star-review}
\end{align}
On the triangular face,
\begin{align}
        T_{0}(Q,D)
        &=
        f_D(e)^3,
        \label{eq:perm-tri-x0-review}\\
        T_{0}^{*}(Q,D)
        &=
        \frac{1}{Q!}
        \sum_{\substack{g_1,g_2,g_3\in S_Q\\g_1g_2g_3=e}}
        f_D(g_1)f_D(g_2)f_D(g_3).
        \label{eq:perm-tri-x0star-review}
\end{align}
Below we evaluate these four quantities for
\(f_D(g)=D^{\cycles(g)}\).  The two expressions in each pair are not
interchangeable: they are the two principal factors obtained by carrying out
the duality analysis in conjunction with the star-triangle transformation at the same finite-basis level.

\subsection{Honeycomb lattice}
On the honeycomb lattice, fix the three external permutations of a Y-shaped
block to the identity and sum the central permutation.  
The original principal factor is
\begin{equation}
        H_0(Q,D)
        =
        \sum_{\sigma\in S_Q}D^{3\cycles(\sigma)}
        =
        \Zcyc(D^3)
        =
        \frac{\Gamma(D^3+Q)}{\Gamma(D^3)} .
        \label{eq:H0}
\end{equation}
After duality, the three dual edge variables of the triangular face are
independent in the principal configuration.  With the same normalization as
in the single-bond duality, the dual principal factor is
\begin{equation}
        H_0^*(Q,D)
        =
        \frac{\Zcyc(D)^3}{Q!}
        =
        \frac{1}{\Gamma(Q+1)}
        \left[
        \frac{\Gamma(D+Q)}{\Gamma(D)}
        \right]^3 .
        \label{eq:H0-dual}
\end{equation}
The honeycomb finite-\(Q\) condition is therefore
\begin{equation}
        \Zcyc(D^3)
        =
        \frac{\Zcyc(D)^3}{Q!}.
        \label{eq:honeycomb-finiteQ}
\end{equation}
Taking the coefficient linear in \(Q\) gives the closed replica-limit
equation
\begin{equation}
        \psi(D^3)=3\psi(D)+\gamma .
        \label{eq:honeycomb-q0}
\end{equation}
Its positive solution is
\begin{equation}
        \Dc^{\hex}=2.634929344884\cdots .
        \label{eq:honeycomb-result}
\end{equation}

\subsection{Triangular lattice}
For the triangular lattice, the original principal factor of the triangular
face is
\begin{equation}
        T_0(Q,D)=D^{3Q}.
        \label{eq:T0}
\end{equation}
The dual face factor contains the constraint around the triangle:
\begin{equation}
        T_0^*(Q,D)
        =
        \Ytri(Q,D)
        =
        \frac{1}{Q!}
        \sum_{\substack{g_1,g_2,g_3\in S_Q\\g_1g_2g_3=e}}
        D^{\cycles(g_1)+\cycles(g_2)+\cycles(g_3)} .
        \label{eq:Y-constrained}
\end{equation}
The calculation is not straightforward.
However, the duality relation for mutually dual non-self-dual lattices
uses the single-bond principal factors.  
For the present permutation model, each edge principal factor is
\begin{equation}
        x_0(Q,D)=D^Q,\qquad
        x_0^*(Q,D)=\frac{\Zcyc(D)}{\sqrt{Q!}} .
        \label{eq:single-bond-x0}
\end{equation}
Therefore, the single-bond mutual-dual relation is
\begin{equation}
        D_\tri^Q D_\hex^Q
        =
        \frac{\Zcyc(D_\tri)\Zcyc(D_\hex)}{Q!}.
        \label{eq:single-bond-pair}
\end{equation}
Taking the coefficient linear in \(Q\) gives
\begin{equation}
        \log D_\tri+\log D_\hex
        =
        \psi(D_\tri)+\psi(D_\hex)+\gamma .
        \label{eq:single-bond-pair-q0}
\end{equation}
Inserting the solution of the honeycomb block equation
\eqref{eq:honeycomb-result} into Eq.~\eqref{eq:single-bond-pair-q0} gives
\begin{equation}
        D_\tri^{\rm pair}=1.475661534848\cdots .
        \label{eq:tri-pair-result}
\end{equation}
This is the estimate of the critical point on the triangular lattice.

As detailed in Appendix~\ref{app:symmetric-group}, the direct evaluation of the triangular constrained sum gives
\begin{equation}
        \Ytri(Q,D)
        =
        \sum_{\lambda\vdash Q}
        \frac{(D)_\lambda^3}{h_\lambda^2},
        \label{eq:Y-young}
\end{equation}
where \((D)_\lambda\) is the content product and \(h_\lambda\) is the hook
product of the Young diagram \(\lambda\).  
If the triangular projection is imposed by itself, the finite-\(Q\) condition is
\begin{equation}
        D^{3Q}=\Ytri(Q,D).
        \label{eq:triangular-finiteQ}
\end{equation}
For finite \(Q\), these equations give estimates of the critical points on the triangular lattice that are consistent with the results obtained by the duality relation from the honeycomb-lattice estimates, as listed in Table~\ref{tab:finiteQ}.

\begin{table}[t]
\centering
\begin{tabular}{c|cc}
\toprule
\(Q\) & \(D_\tri\) 
    & \(D_\hex\) \\
\midrule
2 & \(1.732050807569=\sqrt3\) & \(3.732050807569=2+\sqrt3\)\\
3 & \(1.814949258469\) & \(4.133982205418\)\\
4 & \(1.885773637489\) & \(4.490189417478\)\\
5 & \(1.948340595665\) & \(4.814308190660\)\\
6 & \(2.004824387604\) & \(5.114293715547\)\\
\bottomrule
\end{tabular}
\caption{Finite-\(Q\) results.}
\label{tab:finiteQ}
\end{table}

To take the replica limit we write
\begin{equation}
        \Ytri(Q,D)
        =
        \frac{D^{3Q}}{\Gamma(Q+1)}
        \left[
        1+\sum_{r\ge1}c_r(Q)D^{-2r}
        \right].
        \label{eq:Y-expansion}
\end{equation}
For fixed \(r\), Farahat-Higman stability implies that \(c_r(Q)\) is a
polynomial in \(Q\) \cite{FarahatHigman}.  
We define
\begin{equation}
        b_r=c_r'(0),
        \qquad
        \Sseries(D)=\sum_{r\ge1}\frac{b_r}{D^{2r}} .
        \label{eq:br-and-S}
\end{equation}
Then
\begin{equation}
        \left.
        \partial_Q\log \Ytri(Q,D)
        \right|_{Q=0}
        =
        3\log D+\gamma+\Sseries(D).
        \label{eq:Y-derivative}
\end{equation}
The replica coefficient of the direct triangular block condition
\eqref{eq:triangular-finiteQ} is therefore
\begin{equation}
        \Sseries(D)=-\gamma .
        \label{eq:tri-q0-condition}
\end{equation}
This is an exact algebraic condition obtained from the direct evaluation of the triangular summation.  
The series is asymptotic at the physical value of \(D\).  
The Borel-Pade and Borel-Leroy-Pade analysis of this diagnostic is given in Appendix~\ref{app:pade}.
This gives a nearby diagnostic value, but we do not use it as the critical-point estimate.

\section{Results and Discussion}
\label{sec:discussion}
In the replica limit, we find the honeycomb block value and the associated
single-bond mutual-dual triangular value
\(\Dc^{\tri}=1.475661534848\cdots\) and
\(\Dc^{\hex}=2.634929344884\cdots\).
The location of \(D=2\) is physically special because it is the qubit bond
dimension.  
In random tensor networks and monitored random circuits, the ordering transition of the associated permutation model is interpreted as an entanglement transition separating an area-law, disentangling phase from a volume-law, entangling phase \cite{Skinner2019,Li2019,Zhou2019,Romain2019}.
The volume-law phase is also understood as a quantum-error-correcting or
information-protecting phase: scrambling can protect quantum information
against local measurements or errors below a threshold, whereas the area-law
phase has no such robust channel capacity \cite{Choi2020}.  
Our results should therefore be read as a statement
about the resource character of qubit realizations on the two lattices.  
For the triangular lattice, \(D=2\) lies above the estimated critical bond
dimension, so the qubit model is on the ordered, volume-law side in this
duality estimate.  
On the other hand, for the honeycomb lattice, \(D=2\) lies below the estimated
critical bond dimension, so the qubit model remains on the disordered,
area-law side.  
In this sense, the simple honeycomb geometry is not expected to provide the same volume-law entangled resource, or the same protected quantum information, at qubit bond dimension.  
The difference is consistent with the lower coordination of the honeycomb lattice: fewer local interactions require a larger bond dimension, equivalently a lower effective temperature, before the ordered phase is reached.

The direct triangular star-triangle projection remains useful as a
group-theoretic check.  
It asks for the coefficient \(\Sseries(D)\) in the constrained three-permutation sum \eqref{eq:Y-constrained}; the Farahat-Higman stable expansion and Borel-Pade resummation in Appendix~\ref{app:pade} provide an explicit formula for that triangular block coefficient and yield non-trivial results stemming from the duality analysis rather than from group theory alone.  

\section{Conclusion}
\label{sec:conclusion}

We have performed the duality analysis in conjunction with the star-triangle transformation for the symmetric-group permutation model on the triangular and honeycomb lattices.
The honeycomb side is a Y-block central-permutation
sum and gives a closed replica-limit equation.  
The corresponding mutual-dual relation gives a paired estimate on the triangular lattice.  
Separately, the dual constrained three-permutation face
sum gives an explicit Young-diagram representation of the triangular block
coefficient, whose numerical evaluation is obtained by
Borel-Pade resummation and yields a nearby diagnostic estimate.

The fact that the triangular estimate lies below two while the honeycomb
estimate lies above two has a direct quantum-information interpretation.  
At the qubit bond dimension, the triangular geometry is estimated to lie on the
volume-law, resourceful side of the entanglement transition, whereas the honeycomb geometry is estimated to remain on the area-law side.  
The honeycomb result should therefore be viewed as evidence that this simple
geometry is less favorable for qubit-bond random tensor-network applications.

The success of the duality analysis in conjunction with the star-triangle transformation suggests the possibility of further analysis of critical points in random quantum circuits, whose statistical-mechanical formulations are closely related to random tensor networks.
A natural future direction is the explicit duality analysis of random quantum circuit models, which have corresponding statistical-mechanical models on square, triangular, and honeycomb lattices.

\section*{Acknowledgements}

The author thanks T. Matsuo for fruitful discussions and also acknowledges financial support from the Cross-ministerial Strategic Innovation Promotion Program (SIP) of the Cabinet Office (No. 23836436).

\appendix

\section{Symmetric-group identities}
\label{app:symmetric-group}

The cycle-index identity used in the main text is
\begin{equation}
        \Zcyc(a)
        =
        \sum_{g\in S_Q}a^{\cycles(g)}
        =
        a(a+1)\cdots(a+Q-1)
        =
        \frac{\Gamma(a+Q)}{\Gamma(a)}.
        \label{eq:cycle-index-app}
\end{equation}
Taking the logarithmic derivative with respect to \(Q\) at \(Q=0\) gives
Eq.~\eqref{eq:cycle-index-derivative}.

Let \(\lambda\vdash Q\) label an irreducible representation of \(S_Q\), with
dimension \(d_\lambda\), character \(\chi_\lambda\), and matrix
\(\rho^\lambda(g)\).  The group delta function is
\begin{equation}
        \delta(e,g)
        =
        \frac{1}{Q!}
        \sum_{\lambda\vdash Q}
        d_\lambda\chi_\lambda(g).
        \label{eq:delta}
\end{equation}
For a central function \(f(g)\), the Fourier transform is scalar on each
irreducible representation:
\begin{equation}
        \hat f(\lambda)
        =
        \frac{1}{d_\lambda}
        \sum_{g\in S_Q}f(g)\chi_\lambda(g).
        \label{eq:central-fourier}
\end{equation}
For \(f_D(g)=D^{\cycles(g)}\), the Jucys-Murphy identity gives the content
product \cite{Jucys}
\begin{equation}
        \hat f_D(\lambda)
        =
        (D)_\lambda
        =
        \prod_{\Box\in\lambda}\bigl(D+\cont(\Box)\bigr),
        \label{eq:content-product}
\end{equation}
where \(\cont(\Box)=j-i\) for a box in row \(i\) and column \(j\).
The hook product is denoted by \(h_\lambda\), and
\(d_\lambda=Q!/h_\lambda\).

We now derive the triangular constrained sum.  Starting from
Eq.~\eqref{eq:Y-constrained} and inserting Eq.~\eqref{eq:delta},
\begin{align}
        \Ytri(Q,D)
        &=
        \frac{1}{Q!}
        \sum_{g_1,g_2,g_3}
        f_D(g_1)f_D(g_2)f_D(g_3)
        \delta(e,g_1g_2g_3)
        \nonumber\\
        &=
        \frac{1}{(Q!)^2}
        \sum_{\lambda\vdash Q}
        d_\lambda
        \sum_{g_1,g_2,g_3}
        f_D(g_1)f_D(g_2)f_D(g_3)
        \chi_\lambda(g_1g_2g_3).
        \label{eq:Y-derivation-1}
\end{align}
Since \(f_D\) is central, convolution by \(f_D\) acts on representation
\(\lambda\) as multiplication by \((D)_\lambda\).  Applying this three times
and using \(d_\lambda=Q!/h_\lambda\) gives
\begin{equation}
        \Ytri(Q,D)
        =
        \sum_{\lambda\vdash Q}
        \frac{(D)_\lambda^3}{h_\lambda^2},
        \label{eq:Y-derivation-2}
\end{equation}
which is Eq.~\eqref{eq:Y-young}.

\section{Farahat-Higman coefficients}
\label{app:fh}

After factoring out \(D^{3Q}/\Gamma(Q+1)\) as in
Eq.~\eqref{eq:Y-expansion}, the fixed-order coefficient \(c_r(Q)\) is a
polynomial in \(Q\).  This is the relevant Farahat-Higman stability
statement for the replica calculation \cite{FarahatHigman}.  In practice we
determine this polynomial by evaluating the Young-diagram sum at
sufficiently many positive integers \(Q\) and interpolating.  The coefficient
entering the replica limit is \(b_r=c_r'(0)\).

The first coefficients are
\begin{equation}
\begin{split}
        b_1&=-\frac32,\\
        b_2&=\frac54,\\
        b_3&=-\frac{41}{6},\\
        b_4&=\frac{2751}{40},\\
        b_5&=-\frac{2057}{2}.
\end{split}
        \label{eq:first-br}
\end{equation}
The full list through \(r=24\) used in the numerical analysis is shown in
Appendix~\ref{app:pade}.

\section{Borel-Pade analysis}
\label{app:pade}

The triangular condition is
\begin{equation}
        \gamma+\Sseries(D)=0,
        \qquad
        \Sseries(D)=\sum_{r\ge1}\frac{b_r}{D^{2r}}.
        \label{eq:pade-condition}
\end{equation}
The Borel-Leroy transform is
\begin{equation}
        \mathcal B_\beta(t)
        =
        \sum_{r\ge1}
        \frac{b_r}{\Gamma(r+\beta+1)}t^r,
        \label{eq:borel-leroy}
\end{equation}
and the resummed value is
\begin{equation}
        \Sseries_\beta(D)
        =
        \int_0^\infty
        \e^{-t}t^\beta
        \mathcal B_\beta(t/D^2)\,\dd t.
        \label{eq:borel-integral}
\end{equation}
In the numerical analysis \(\mathcal B_\beta\) is replaced by a Pade
approximant.  We keep near-diagonal approximants and discard those with
poles on the positive Borel axis \cite{Borel,Pade}.

At ordinary Borel parameter \(\beta=0\), representative accepted
near-diagonal approximants give
\begin{center}
\begin{tabular}{c|c}
\toprule
Approximant & root of \(\gamma+\Sseries(D)=0\)\\
\midrule
\([10/7]\) & 1.484593866\\
\([8/9]\) & 1.484610587\\
\([11/8]\) & 1.484592472\\
\([9/10]\) & 1.484606991\\
\([11/9]\) & 1.484612874\\
\([10/10]\) & 1.484618781\\
\([13/11]\) & 1.484635889\\
\([12/12]\) & 1.484643831\\
\bottomrule
\end{tabular}
\end{center}
Borel-Leroy variations with \(\beta\) between \(0\) and \(10\) give the same
stable window once approximants with positive-axis poles and clear
off-diagonal outliers are discarded.  This gives the following direct
triangular block diagnostic,
\begin{equation}
        D_\tri^{\rm direct}=1.48463(10).
        \label{eq:pade-quoted}
\end{equation}

The coefficients \(b_r\) are
\begin{center}
\begin{tabular}{c|l}
\toprule
\(r\) & \(b_r\)\\
\midrule
1 & \(-3/2\)\\
2 & \(5/4\)\\
3 & \(-41/6\)\\
4 & \(2751/40\)\\
5 & \(-2057/2\)\\
6 & \(8652653/420\)\\
7 & \(-1037115/2\)\\
8 & \(1258318603/80\)\\
9 & \(-70333566323/126\)\\
10 & \(4988697711069/220\)\\
11 & \(-2073563173943/2\)\\
12 & \(574848226265015081/10920\)\\
13 & \(-5871772004472249/2\)\\
14 & \(3564352044109610437/20\)\\
15 & \(-5398116381150146790965/462\)\\
16 & \(2235381004750746734965611/2720\)\\
17 & \(-123304305705128771759525/2\)\\
18 & \(305437421844186365590172361079/62244\)\\
19 & \(-824867594394668052628295919/2\)\\
20 & \(80179505221780184903290437293081/2200\)\\
21 & \(-141629125684800068198012245049861/42\)\\
22 & \(149688872479132067915733225978457647/460\)\\
23 & \(-65248235249478356448876078382617275/2\)\\
24 & \(1256628358148640057766935797821738630092841/371280\)\\
\bottomrule
\end{tabular}
\end{center}


\begin{thebibliography}{99}

\bibitem{Kramers1941}
H.~A. Kramers and G.~H. Wannier,
Statistics of the Two-Dimensional Ferromagnet. Part I,
Phys. Rev. \textbf{60}, 252 (1941).

\bibitem{WuWang1976}
F.~Y. Wu and Y.~K. Wang,
Duality transformation in a many-component spin model,
J. Math. Phys. \textbf{17}, 439 (1976).

\bibitem{Kogut1979}
J.~B. Kogut,
An introduction to lattice gauge theory and spin systems,
Rev. Mod. Phys. \textbf{51}, 659 (1979).

\bibitem{Nishimori1979}
H.~Nishimori,
Conjecture on the exact transition point of the random Ising ferromagnet,
J. Phys. C \textbf{12}, L905 (1979).

\bibitem{Nishimori2002}
H.~Nishimori and K.~Nemoto,
Duality and multicritical point of two-dimensional spin glasses,
J. Phys. Soc. Jpn. \textbf{71}, 1198 (2002);
\href{https://arxiv.org/abs/cond-mat/0111354}{arXiv:cond-mat/0111354}.

\bibitem{Maillard2003}
J.-M. Maillard, K. Nemoto, and H. Nishimori,
Symmetry, complexity and multicritical point of the two-dimensional spin
glass,
J. Phys. A \textbf{36}, 9799 (2003);
\href{https://arxiv.org/abs/cond-mat/0306154}{arXiv:cond-mat/0306154}.

\bibitem{Nishimori2006}
H.~Nishimori and M.~Ohzeki,
Location of the multicritical point for the Ising spin glass on the
triangular and hexagonal lattices,
J. Phys. Soc. Jpn. \textbf{75}, 034004 (2006);
\href{https://arxiv.org/abs/cond-mat/0601356}{arXiv:cond-mat/0601356}.

\bibitem{Ohzeki2008hl}
M.~Ohzeki, H.~Nishimori, and A.~N. Berker,
Multicritical points for the spin glass models on hierarchical lattices,
Phys. Rev. E \textbf{77}, 061116 (2008);
\href{https://arxiv.org/abs/0802.2760}{arXiv:0802.2760}.

\bibitem{Ohzeki2009}
M.~Ohzeki,
Locations of multicritical points for spin glasses on regular lattices,
Phys. Rev. E \textbf{79}, 021129 (2009);
\href{https://arxiv.org/abs/0811.0464}{arXiv:0811.0464}.

\bibitem{Ohzeki2011slope}
M.~Ohzeki, C.~K. Thomas, H.~G. Katzgraber, H.~Bombin, and M.~A. Martin-Delgado,
Universality in phase boundary slopes for spin glasses on self-dual lattices,
J. Stat. Mech. \textbf{2011}, P02004 (2011).

\bibitem{Ohzeki2015}
M.~Ohzeki and J.~L. Jacobsen,
High-precision phase diagram of spin glasses from duality analysis with
real-space renormalization and graph polynomials,
J. Phys. A \textbf{48}, 095001 (2015);
\href{https://arxiv.org/abs/1410.0166}{arXiv:1410.0166}.

\bibitem{Zhou2019}
T.~Zhou and A.~Nahum,
Emergent statistical mechanics of entanglement in random unitary circuits,
Phys. Rev. B \textbf{99}, 174205 (2019);
\href{https://arxiv.org/abs/1804.09737}{arXiv:1804.09737}.

\bibitem{Romain2019}
R.~Vasseur, A.~C. Potter, Y.-Z. You, and A.~W.~W. Ludwig,
Entanglement transitions from holographic random tensor networks,
Phys. Rev. B \textbf{100}, 134203 (2019);
\href{https://arxiv.org/abs/1807.07082}{arXiv:1807.07082}.

\bibitem{Yimu2020}
Y.~Bao, S.~Choi, and E.~Altman,
Theory of the phase transition in random unitary circuits with measurements,
Phys. Rev. B \textbf{101}, 104301 (2020).

\bibitem{Jian2020}
C.-M. Jian, Y.-Z. You, R.~Vasseur, and A.~W.~W. Ludwig,
Measurement-induced criticality in random quantum circuits,
Phys. Rev. B \textbf{101}, 104302 (2020).

\bibitem{Hayden2016}
P.~Hayden, S.~Nezami, X.-L. Qi, N.~Thomas, M.~Walter, and Z.~Yang,
Holographic duality from random tensor networks,
J. High Energy Phys. \textbf{2016}, 9 (2016);
\href{https://arxiv.org/abs/1601.01694}{arXiv:1601.01694}.

\bibitem{Drouffe1979}
J.~M. Drouffe, C.~Itzykson, and J.~B. Zuber,
Lattice models with a solvable symmetry group,
Nucl. Phys. B \textbf{147}, 132 (1979).

\bibitem{Drouffe1978}
J.~M. Drouffe,
Transitions and duality in gauge lattice systems,
Phys. Rev. D \textbf{18}, 1174 (1978).

\bibitem{Buchstaber2003}
V.~M. Buchstaber and M.~I. Monastyrsky,
Generalized Kramers-Wannier duality for spin systems with non-commutative
symmetry,
J. Phys. A \textbf{36}, 7679 (2003).

\bibitem{PermutationDuality}
M.~Ohzeki,
Duality analysis in symmetric group and its application to random tensor
network model,
\href{https://arxiv.org/abs/2310.14140}{arXiv:2310.14140}.

\bibitem{Ito2026}
R.~Ito, T.~Matsuo, and M.~Ohzeki,
Analysis of critical points in a permutation model on hierarchical lattices
by real-space renormalization group,
\href{https://arxiv.org/abs/2605.25683}{arXiv:2605.25683}.

\bibitem{BaxterBook}
R.~J. Baxter,
\textit{Exactly Solved Models in Statistical Mechanics},
Academic Press, London (1982).

\bibitem{Takeda2005}
K.~Takeda, T.~Sasamoto, and H.~Nishimori,
Exact location of the multicritical point for finite-dimensional spin
glasses: A conjecture,
J. Phys. A \textbf{38}, 3751 (2005);
\href{https://arxiv.org/abs/cond-mat/0501372}{arXiv:cond-mat/0501372}.

\bibitem{FarahatHigman}
H.~K. Farahat and G.~Higman,
The centres of symmetric group rings,
Proc. R. Soc. Lond. A \textbf{250}, 212 (1959).

\bibitem{Skinner2019}
B.~Skinner, J.~Ruhman, and A.~Nahum,
Measurement-induced phase transitions in the dynamics of entanglement,
Phys. Rev. X \textbf{9}, 031009 (2019);
\href{https://arxiv.org/abs/1808.05953}{arXiv:1808.05953}.

\bibitem{Li2019}
Y.~Li, X.~Chen, and M.~P.~A. Fisher,
Measurement-driven entanglement transition in hybrid quantum circuits,
Phys. Rev. B \textbf{100}, 134306 (2019);
\href{https://arxiv.org/abs/1901.08092}{arXiv:1901.08092}.

\bibitem{Choi2020}
S.~Choi, Y.~Bao, X.-L. Qi, and E.~Altman,
Quantum error correction in scrambling dynamics and measurement-induced phase
transition,
Phys. Rev. Lett. \textbf{125}, 030505 (2020);
\href{https://arxiv.org/abs/1903.05124}{arXiv:1903.05124}.

\bibitem{Jucys}
A.~A. Jucys,
Symmetric polynomials and the center of the symmetric group ring,
Rep. Math. Phys. \textbf{5}, 107 (1974).

\bibitem{Borel}
E.~Borel,
Memoire sur les series divergentes,
Ann. Sci. Ecole Norm. Sup. \textbf{16}, 9 (1899).

\bibitem{Pade}
H.~Pade,
Sur la representation approchee d'une fonction par des fractions
rationnelles,
Ann. Sci. Ecole Norm. Sup. \textbf{9}, 3 (1892).

\end{thebibliography}
\end{document}